\documentclass[final,3p,times]{elsarticle}
\usepackage[utf8]{inputenc}      
\usepackage[LGR,T1]{fontenc}     
\usepackage{graphicx}
\usepackage{amssymb}
\usepackage{amsmath}             
\usepackage{multirow}            
\usepackage[hidelinks]{hyperref} 

\newcounter{bla}

\journal{Solid State Sciences}

\begin{document}

\begin{frontmatter}

\title{The structure of thin boron nanowires predicted using evolutionary computations}
\author[a]{Tomasz Tarkowski}
\author[b]{Nevill Gonzalez Szwacki\corref{author}}
\cortext[author] {Corresponding author.\\\textit{E-mail address:} gonz@fuw.edu.pl}
\address[a]{Institute of Theoretical Physics, Faculty of Physics, University of Warsaw, Pasteura 5, PL-02093 Warszawa, Poland}
\address[b]{Institute of Experimental Physics, Faculty of Physics, University of Warsaw, Pasteura 5, PL-02093 Warszawa, Poland}

\begin{abstract}
This work describes the implementation of a genetic algorithm-based strategy combined with first-principles computations for identifying the structure of the most stable boron 1D structures. We focus our attention on the structure of ultrathin 1D boron structures given the lack of previous experimental and theoretical work on this topic. Our methodology yields reasonable structural candidates for further optimizations at the DFT level with tighter convergence criteria. The simulations involved 1D structures with up to 8 atoms per unit cell. We have identified four main groups of structures: flat nanowires (monatomic-height stripes) with triangular or triangular and “square” motifs,
stripes with larger holes, nanowires with an open tubular shape, and regular nanowires. The diameter-dependent structural changes are discussed. 

\end{abstract}


\end{frontmatter}

\section{Introduction}
Genetic algorithms (GAs) serve as a framework for solving optimization problems. One such problem is crystal structure prediction (CSP) based solely on stoichiometry. Brute force CSP is prohibitively expensive \cite{Oganov2006} and often lacks proper solution \cite{Oganov2006,Maddox1988}.

Boron is one of the least understood chemical elements in the periodic table \cite{GonzalezSzwacki2021}. It was discovered in 1808 \cite{Davy1809} and first obtained in pure form in 1909 \cite{Weintraub1910}. Ultrapure boron single crystals for the semiconductor industry are obtained from the decomposition of diborane assisted with Czochralski or zone melting purification processes \cite{Starks1960}. Boron has two naturally occurring isotopes, $^{10}{\rm B}$ and quantitatively dominant $^{11}{\rm B}$ (both stable) \cite{NISTBoron}, and over a dozen radioisotopes up to $^{21}{\rm B}$ \cite{Leblond2018}, which in this particular example decays through a two-neutron emission channel. Boron in the Universe and on the Earth comes from the spallation process \cite{Meneguzzi1971}. Its abundance in the continental crust is at the level of $0{.}001\%$, which ranks this element at position number 37, after lead and before thorium \cite{Haynes2016}. Boron is important to life on Earth---it co-creates plant cell walls \cite{Matoh1997}. This element was discovered on Mars in Gale Crater in way suggesting that water was there in the past \cite{Gasda2017}. On Earth, on the west edge of the Mojave Desert (USA), in a census-designated place with not accidental name of Boron (California, USA) is located the greatest borax (${\rm Na}_2 {\rm B}_4 {\rm O}_7 \!\cdot\! 10 {\rm H}_2 {\rm O}$) mine on the planet. Borax can be easily converted into boric acid and then into boron trioxide, and finally into elemental boron.

Boron has an electron configuration of $[{\rm He}]\ 2s^2\ 2p^1$, which means that it cannot form carbon-like hybridization. Instead, it can form e.g. covalent bonds with three centers sharing two electrons (3c-2e) \cite{Calabrese1971}. All known three dimensional (3D) boron allotropes are based on 12-atom icosahedra ${\rm B}_{12}$ \cite{Delaplane1988a,Delaplane1988b,Saeed2021}.

Similarly to carbon, boron can form low-dimensional structures. In the two-dimensional (2D) case, theoretically, the most stable neutral structure is the $\alpha$-sheet \cite{Tang2007}, however in electron-rich environments, e.g., during production on metallic substrates, the negatively charged $\delta$-sheet would be preferred \cite{Tarkowski2018}. This structure was obtained experimentally on a silver substrate by molecular beam epitaxy (MBE) \cite{Feng2016}. The first 2D boron structures were obtained in 2015 chemical vapor deposition (CVD) \cite{Tai2015} and MBE \cite{Mannix2015}. 2D boron production is now standard procedure \cite{Ranjan2020,Xie2020,Hou2020}.

In the one-dimensional (1D) case, boron forms nanowires \cite{Otten2002,Wang2003,Ding2006,Tian2008,Patel2015} and nanotubes \cite{Patel2015,Gindulyte1998,Ciuparu2004,Kunstmann2014,Kwun2018}. 1D boron of the BDC (\emph{boron double chain}) type (cf. $s_4$ structure from Ref.~\cite{Tarkowski2021}) is metallic and so robust that it does not undergo Peierls transition \cite{Peierls1985} whereas it has a Kohn anomaly in the phonon dispersion relation \cite{Kohn1959}. 1D boron of the BTC (\emph{boron triple chain}) type is energetically more favorable than the BDC type \cite{Tarkowski2021}.

Zero-dimensional (0D) boron forms structures similar to carbon fullerenes, e.g., ${\rm B}_{80}$ (\emph{buckyball}) \cite{GonzalezSzwacki2007a} and ${\rm B}_{40}$ (\emph{borospherene}) molecules \cite{Zhai2014}. Experimental and theoretical research on ${\rm B}_{36}$ cluster \cite{Piazza2014} formed the basis for obtaining 2D boron structures.

Boron can form materials with other chemical elements (alloys). One such example is boron nitride nanotube (BNNT) \cite{Rubio1994,Chopra1995}. Paper made of BNNTs, contrary to ordinary paper or made of carbon nanotunes, is flame resistant \cite{Kim2015}. BNNTs can also spontaneously reconstruct after bending \cite{Goldberg2009}.

Scientific and engineering breakthroughs of the next two decades, especially the increase in computing power, enable---at least partially---the realization of CSP. At the beginning of this century, Oganov et al. \cite{Oganov2006} realized a computer program called USPEX, which merges the GA approach with density functional theory (DFT) calculations. This method, despite being computationally quite expensive, is fruitful---boron allotrope $\gamma$-${\rm B}_{28}$ \cite{Oganov2009,Solozhenko2008} and high-pressure sodium \cite{Ma2009} are two examples where numerical simulations were experimentally confirmed.

The USPEX program is not the first case of an evolutionary computation (EC) software package applied to the structure prediction task, a similar approach was previously used for the protein folding problem \cite{Unger1993}. On the other hand, it seems that Oganov's work was the first attempt which encouraged the condensed matter physics community for this method---sometime later Zurek et al. \cite{Lonie2011a} presented the XtalOpt program that is using similar GA/DFT combination, while scientists working on the CALYPSO project \cite{Wang2012,Tong2019} implemented the particle swarm optimization (PSO) algorithm to determine the crystal structure of materials.

Boron in its 1D form remains the least studied, both theoretically and experimentally \cite{ GonzalezSzwacki2021,Kondo2017}. In this work, we find the structure of boron nanowires with diameters smaller than 1~nm. We are interested to study the structural changes of these ultrathin nanowires as a function of their diameter. To achieve this goal, the most stable forms of all-boron nanowires are found using a developed for this purpose software package called Quil\"e \cite{tarkowski_tomasz_2022_6484743}.

\section{Theoretical approach}
An evolutionary algorithm is a non-deterministic optimization method that in each run can generate different solutions, even if the initial structures are the same. The task to be solved in this work is to find---with the use of GA and DFT---the optimal crystal structure of monoatomic nanowires. Two groups of 1D structures will be considered. In the first group, the boron structures are confined to planarity. This kind of structures are commonly known as atomic stripes. In the second group of structures, the out of plane position of atoms is allowed and as a result of the structure optimization we can end up having buckled stripes, nanowires, and even nanotubes. The first step in our nanowire CSP's problem is to define the unit cell in terms of predicate formalism.

\subsection{Unit cell definition in terms of predicate $Q$}

The unit cell of our 1D structure is defined in terms of predicate $Q$. Without loss of generality one can make the following assumptions:
\begin{itemize}
\item the 1D structures are periodic in the $z$ direction,
\item the length of the unit cell in the $z$ direction is equal to $h$,
\item atomic positions in the unit cell are described with a sequence of vectors $( \vec{r}_i )_{i=0}^{N_{\rm a}-1}$ fulfilling the condition $\forall i, j \in \iota_{N_{\rm a}} \colon ( i < j \Rightarrow z_i \leq z_j ) \wedge ( i = j \Leftrightarrow \vec{r}_i = \vec{r}_j )$, where $N_{\rm a}$ is the number of atoms in the unit cell,
\item the first atom in the unit cell is at the origin of the coordinate system, i.e., $\vec{r}_0 = \vec{0}$,
\item flat structures are described using Cartesian coordinates $(0,y,z)$, while nanowires are described using cylindrical coordinates $(\rho, \phi, z)$, where we choose $\phi_1 = 0$.
\end{itemize}
It can be easily seen that for flat structures $N_{\rm a} \geq 1$, while for nanowires $N_{\rm a} \geq 3$.

The $Q$ predicate should state two things ($Q \Leftrightarrow Q_0 \wedge Q_1$). Firstly, no two atoms should be too close to each other ($Q_0$ predicate). Secondly, for each atomic pair, the path of interatomic bonds connecting this pair should exist ($Q_1$ predicate). In order to fulfill periodic boundary conditions both predicates (i.e. $Q_0$ and $Q_1$) should be formulated based on two adjacent unit cells. These conditions demand the use of two parameters, labeled as $d_{\min}$ and $d_{\max}$, describing the allowed range of interatomic distances, beyond which the bond cannot be formed because of too small or too big interatomic distance, respectively. The $d_{\min}$ and $d_{\max}$ parameters should be chosen for each material separately and \emph{a priori} knowledge of the chemistry of the material might be helpful to constrain the permitted range of distances.

The $Q_0$ predicate can be formulated as follows:
\begin{eqnarray}
  Q_0 \equiv \forall i, j \in \iota_{N_{\rm a}} \colon & \left( i \neq j \Rightarrow \left| \vec{r}_i - \vec{r}_j \right| \geq d_{\min} \right) \wedge \left( \left| \vec{r}_i - \left( \vec{r}_j + h\hat{e}_z \right) \right| \geq d_{\min} \right) \nonumber \\
  \Leftrightarrow \forall i, j \in \iota_{N_{\rm a}} \colon & \left[ i \neq j \Rightarrow \left( \left| \vec{r}_i - \vec{r}_j \right| \geq d_{\min} \wedge \left| \vec{r}_i - \left( \vec{r}_j + h\hat{e}_z \right) \right| \geq d_{\min} \right) \right] \wedge h \geq d_{\min}
.\end{eqnarray}

To formulate the $Q_1$ predicate, elements of graph theory will be used. Let $V$ be the set of \emph{vertices} consisting of atomic positions taken from two adjacent unit cells, i.e., $V = \{ \vec{r}_i, \vec{r}_i + h\hat{e}_z \mid i \in \iota_{N_{\rm a}} \}$. Let $E$ be the set of \emph{edges} (undirected), which vertices fulfill the condition of maximum distance $d_{\max}$: $E = \{ \{ \vec{q}, \vec{r} \} \subseteq V \mid \vec{q} \neq \vec{r} \wedge | \vec{q} - \vec{r} | \leq d_{\max} \}$. By taking a pair of both objects one obtains \emph{graph} $H = (V, E)$. The $Q_1$ predicate might be now formulated as $Q_1 \equiv \omega (H) = 1$, where $\omega (H)$ is the number of \emph{connected components} of graph $H$, i.e., as a statement that graph $H$ is connected. Graph connectivity can be checked with depth- or breadth-first search, or by analyzing the adjacency matrix. It should be noted that the aforementioned $Q_1$ predicate should be treated as a first approximation of the solution, i.e., every nanowire representation fulfilling $Q_1$ predicate is proper (nanowire is connected), but there are some proper nanowire representations, for which $Q_1$ predicate does not hold (``false negative''). This approximation of $Q_1$ is computationally cheaper than its exact version with checks employing translations along the $z$ axis.

\subsection{Encoding of the unit cell traits. Fitness function}

\begin{figure}
  \centering
  \includegraphics{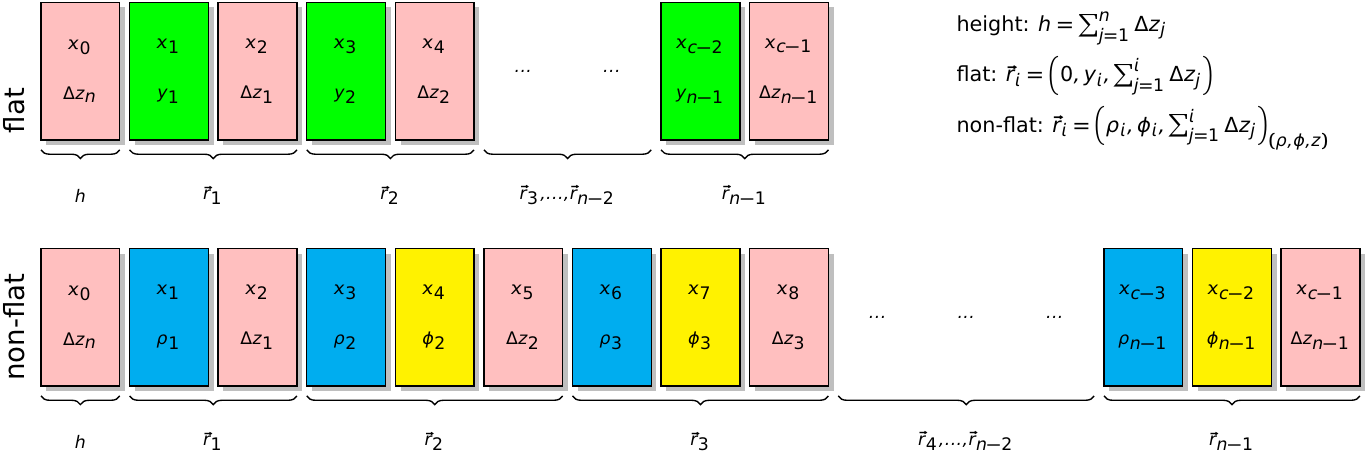}
  \caption{Genotypes for flat and non-flat nanowires with encoded phenotypic traits. The $x_j$ gene can encode the $\vec{r}_i$ coordinate of atom number $i$ or unit cell height $h$. The unit cell height is encoded by multiple genes (polygene). Single gene $x_2$ defines multiple phenotypic traits, i.e., $z_i$ coordinates of all atoms for $i > 0$ and height $h$ (pleiotropy).}
  \label{fig4}
\end{figure}

The information about the phenotype, i.e., atomic positions and unit cell height, should be somehow encoded in the genotype. For this purpose a floating-point representation in the form of $g = (x_i)_{i = 0}^{c - 1} \in [0, 1]^c \subset \mathbb{R}^c$ is used. The genotypes for flat and non-flat cases are shown in Fig.~\ref{fig4}, where the scheme of encoding between the phenotype and genotype is sketched.

One can notice that the genotype length, $c$, depends on the nanowire type:
\begin{equation}
c = \left\{
\begin{array}{ll}
  2 N_{\rm a} - 1 & {\rm flat}, N_{\rm a} \geq 1 \\
  3 (N_{\rm a} - 1) & {\rm non-flat}, N_{\rm a} \geq 3 \\
\end{array}
\right.
.\end{equation}
The same is for the unit cell height, $h$, and atomic positions $(\vec{r}_i)_{i=0}^{N_{\rm a}-1}$. For flat nanowires $h = d_{\max} ( x_0 + \sum_{j=1}^{N_{\rm a} - 1} x_{2j} )$ and the atomic positions for $i \geq 1$ in Cartesian coordinates are:
\begin{equation}
\vec{r}_i = d_{\max} \left( 0 , (N_{\rm a} - 1) \left( \frac{1}{2} x_{2i-1}  - 1 \right) , \sum_{j=1}^i x_{2j} \right)
.\end{equation}
For non-flat nanowires $h = d_{\max} ( x_0 + \sum_{j=1}^{N_{\rm a} - 1} x_{3j-1} )$ and the atomic positions for $i \geq 1$ in cylindrical coordinates are:
\begin{equation}
\vec{r}_i = d_{\max}
\left(
(N_{\rm a} - 1) \cdot \left\{
\begin{array}{l}
  x_1 \\
  x_{3i-3} \\
\end{array}
\right.
,\,
\frac{2\pi}{d_{\max}} \cdot \left\{
\begin{array}{l}
  0 \\
  x_{3i-2} \\
\end{array}
\right.
,\,
\sum_{j = 1}^{i} x_{3j-1}
\right)_{(\rho , \phi , z)}
\left\{
\begin{array}{l}
  i = 1 \\
  i \geq 2 \\
\end{array}
\right.
.\end{equation}

The optimization problem is to find such unit cell height and atomic positions which minimize the total energy. This energy is obtained from self-consistent DFT calculations using the \textsc{Quantum ESPRESSO} (QE) software package \cite{Giannozzi2009, Giannozzi2017}.

\subsection{The case of boron nanowires}

The self-consistent DFT calculations are performed using the projector augmented wave (PAW) technique and the Perdew–Burke–Ernzerhof (PBE) exchange-correlation functional. A Monkhorst–Pack Brillouin zone sampling with a $1 \times 1 \times 8$ $k$-point mesh is used with a half-step offset in the $z$ direction. Broyden electron density mixing with mixing parameter $\beta = 0{.}7$ and a Methfessel-Paxton method with a smearing width of $0{.}01\,{\rm Ry}$ for the electron smearing are used. The kinetic energy cut-off of the plane wave basis is set to $60\,{\rm Ry}$. The PAW pseudopotential is taken from the QE library \cite{B.pbe-n-kjpaw_psl.1.0.0.UPF}. The projection of the distance between extreme atoms of two adjacent unit cells on $x$ and $y$ axes (\emph{buffer thickness}) is equal to $10\,\text{\AA}$, which should be enough to minimize the effects of nanowire replicas placed periodically in the $xy$-plane.

For the purpose of random crystal generation \cite{Tom2020} at the first GA iteration (first generation of genotypes), a mechanism of drawing genes' values from a uniform distribution is used. The genotypes violating the $Q$ predicate are rejected and the drawing procedure is repeated until $Q$ is fulfilled.  Single arithmetic recombination with a random-reset mutation with parameter $p = 1 / c$ applied stochastically is used with recombination and mutation probabilities equal to $p_{\rm r} = 1$ and $p_{\rm m} = 0{.}5$, respectively. The generation size $\mu$ is set to $100$, while the parents' multiset size is set to $2k = 64$. The stochastic universal sampling (SUS) mechanism with linear ranking selection (lin-RS, $s = 2$) is used for parents' selection and selection to the next generation. The genetic search is finished after reaching the fitness function \emph{plateau}, or---more precisely---if after $10$ last generations the fitness function maximum has not improved more than $\Delta E$ with respect to the whole process. The $\Delta E$ value depends on the number of atoms in the unit cell: $\Delta E = \Delta E_{\rm b} \cdot N_{\rm a}$, where $\Delta E_{\rm b} = 1\,{\rm meV} / {\rm atom}$ is a DFT precision of binding energy per atom calculations resulting from the assumed approximations \cite{Hoffman2008}. The bond lengths are constrained between $d_{\min} = 1{.}54\,\text{\AA}$ and $d_{\max} = 2{.}10\,\text{\AA}$ inclusively---these values arise from statistical analysis of known boron structures \cite{Mierzwa2020}.

A complete implementation of boron nanowires CSP is available as an example program in the directory \emph{examples/evenstar/} of the Quil\"e repository. Numerical simulations for flat and non-flat nanowires for $N_{\rm a}$ up to $8$ are performed with the above-mentioned recipe. In order to avoid potential non-full effectiveness of the genetic search, calculations for each $N_{\rm a}$ case, for both flat and non-flat nanowires, are performed $5$ times and calculation logs are available online \cite{CalculationLogs} (please see \emph{evolution/} directory). 

The structures obtained from our GA search are further optimized using QE. Only one or, in case of degeneracy into the same energy, several best structures are further proceeded. For each of these structures, we have optimized the atomic positions and the lattice constants with an increase to a $1 \times 1 \times 16$ $k$-point grid and an increase to a $15\,\text{\AA}$ buffer thickness. The total energy ($E_{\rm tot,\ isolated\ B\ atom}$) for an isolated boron atom is obtained with calculations including spin polarization---boron has an unpaired electron on valence shell (results of calculations are available online \cite{CalculationLogs} in the \emph{single\_atom/} directory). The results of simulations are visualized with the use of the XCrySDen program \cite{Kokalj1999}. Animated visualization of nanowires is presented online \cite{CalculationLogs} in the \emph{results/} directory, where all structures are presented, both before and after relaxation.  

\begin{figure}
  \centering
  \includegraphics[width=\textwidth]{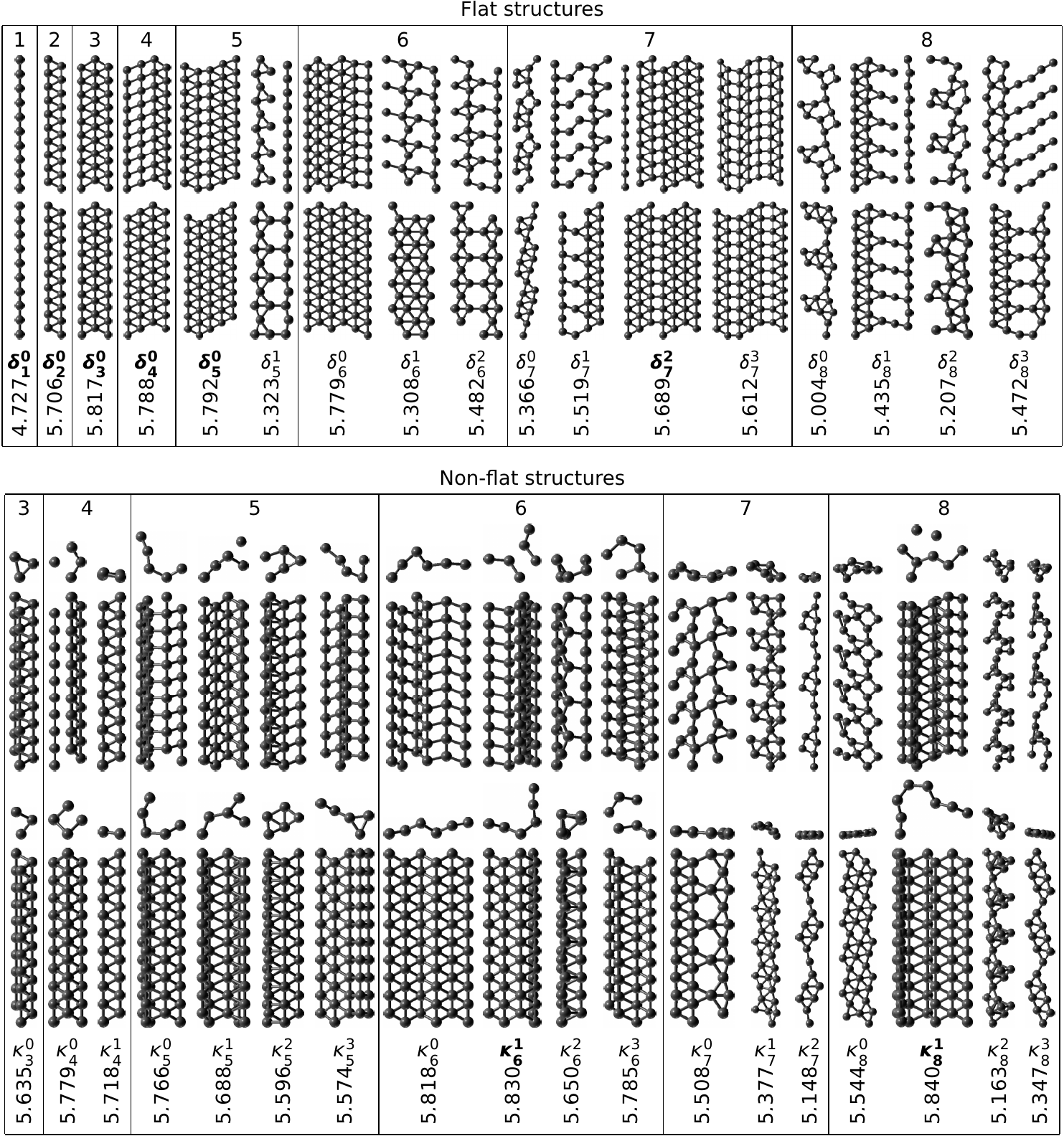}
  \caption{The structure of flat and non-flat nanowires obtained by evolutionary computations followed by DFT relaxation. In both cases, the following data is shown (from top): number of atoms in the unit cell $N_{\rm a}$, evolutionary found structure, the consequently DFT-relaxed structure, the structure label (symbols $\delta_{N_{\rm a}}^i$ or $\kappa_{N_{\rm a}}^i$ for $i \in \mathbb{N}$), and the binding energy per atom, $E_{\rm b}$ (in units of ${\rm eV} / {\rm atom}$), for the DFT-relaxed nanowires. Structures with labels in bold have the highest $E_{\rm b}$ from all nanowires with a given number of atoms in the unit cell. For the non-flat nanowires, beside the side view (alongside the $z$ axis), the top view is shown preserving the scale. Note that the scale between different structures is not preserved.}
  \label{fig5}
\end{figure}

\section{Results and discussion}

The collection of the evolutionary found and DFT-relaxed 1D structures is shown in Fig.~\ref{fig5}. They are divided into two groups: nanowires obtained through GA-evolution of flat $\delta_{N_{\textnormal{a}}}^i$ (gr. \begingroup\fontencoding{LGR}\selectfont διαμέρισμα \endgroup -- flat) and non-flat $\kappa_{N_{\textnormal{a}}}^i$ (gr. \begingroup\fontencoding{LGR}\selectfont κυρτός \endgroup -- convex) structures, where $i \in \mathbb{N}$ is an index labeling structures ($\delta_{N_{\textnormal{a}}}^*$ or $\kappa_{N_{\textnormal{a}}}^*$) having the same number of atoms in the unit cell. The DFT-relaxed structures are presented in Fig.~\ref{fig5} below the corresponding evolutionary found counterparts. The most energetically favorable nanowires within each group ($\delta_{N_{\textnormal{a}}}^*$ and $\kappa_{N_{\textnormal{a}}}^*$) are  highlighted in the figure using bold labels. The binding energy, $E_{\rm b}$, is calculated using the formula:
\begin{equation}
  E_{\rm b} = E_{\rm tot,\ isolated\ B\ atom} - \frac{E_{\rm tot,\ nanowire}}{N_{\rm a}}.
\end{equation}

Analyzing the results presented in Fig.~\ref{fig5}, we can identify two groups of flat structures: stripes with triangular or triangular and “square” motifs, and stripes with larger holes. Looking at non-flat structures, we can find, among others, nanowires with an open tubular shape (e.g. $\kappa_5^0$, $\kappa_6^1$, or $\kappa_8^1$) and regular nanowires (e.g. $\kappa_4^0$, $\kappa_5^2$, or $\kappa_6^2$). The $\delta_2^0$ and $\kappa_4^1$ stripes represent the same structure. Doubling the unit cell size (in the $z$ direction) does not result in any structural changes what confirms the robustness of the BDC structure against Peierls distortions. The $\delta_3^0$ and $\delta_4^0$ stripes are also found in the $\kappa_3^*$ (and also $\delta_6^*$) and $\kappa_4^*$ (and also $\delta_8^*$) series of structures, respectively, and are shown only once in Fig.~\ref{fig5}. The widest considered 1D structure is $\delta_7^3$ (9~{\AA} in diameter) and the thinnest is the boron chain $\delta_1^0$.

\begin{figure}
  \centering
  \includegraphics[width=\textwidth]{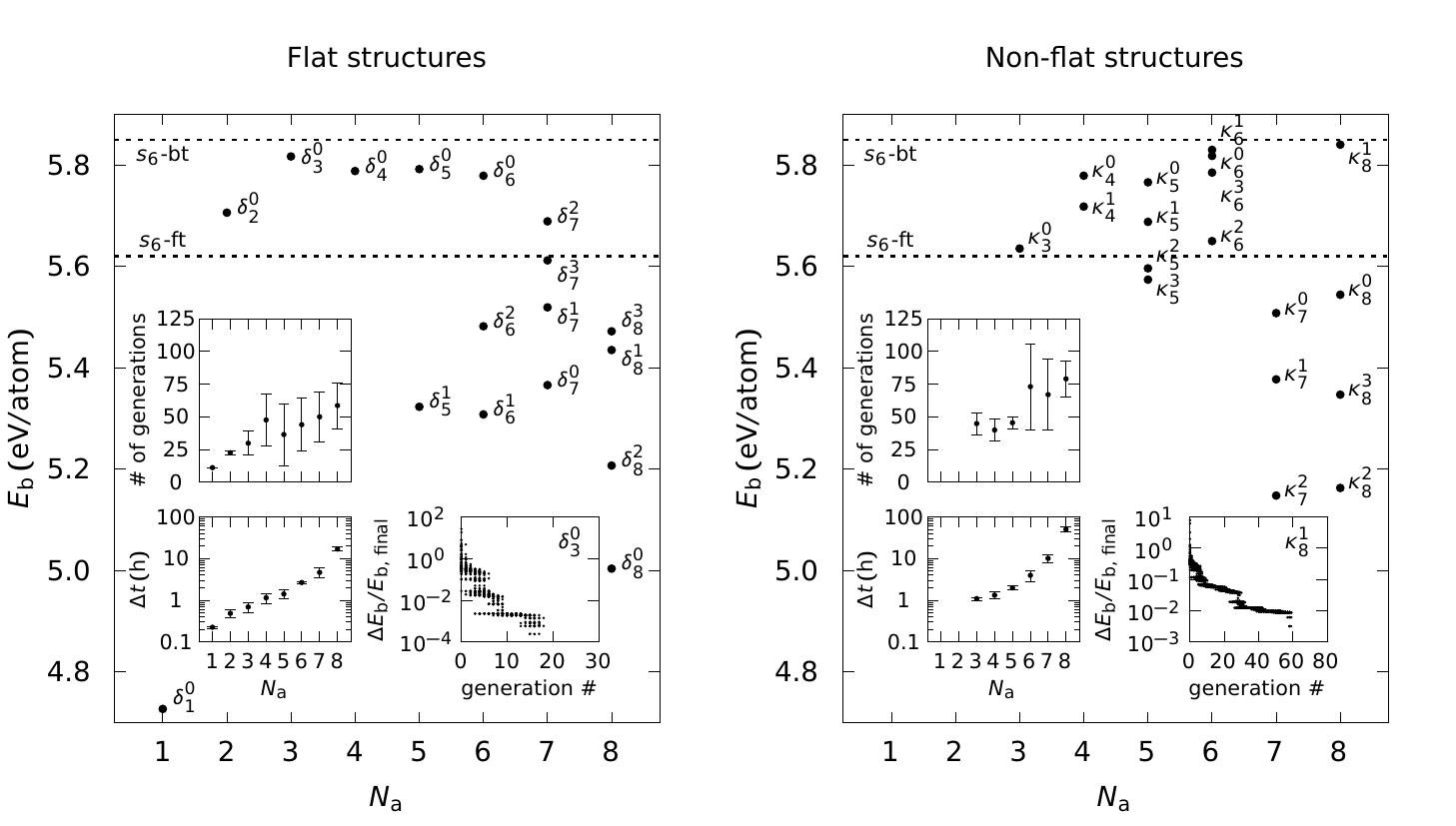}
  \caption{Binding energy per atom, $E_{\rm b}$, for evolutionary obtained and then DFT-relaxed flat (left panel) and non-flat (right panel) nanowires from Fig.~\ref{fig5} as a function of the number of atoms in the unit cell, $N_{\rm a}$. The values of $E_{\rm b}$ for two-dimensional triangular structures, flat $s_6$-ft and buckled $s_6$-bt, marked with dashed lines are taken from \cite{GonzalezSzwacki2007b}. Each panel includes three insets presenting: an average number of generations across 5 computational series for each $N_{\rm a}$ value (top inset), the respective average computational time of the GA-evolution on a single CPU \cite{CPU} machine (bottom left inset), and normalized energy difference $\Delta E_{\rm b} / E_{\rm b,\ final}$, where $\Delta E_{\rm b} = E_{\rm b,\ final} - E_{\rm b}$, as a function of the generation number (bottom right inset) during GA-evolution for the most energetically favorable structures, $\delta_3^0$ and $\kappa_8^1$.}
  \label{fig6}
\end{figure}

From small-diameter structures, the most stable are fully planar stripes with boron triangular motifs. Among all flat structures, the largest value of $E_{\rm b}$ belongs to the $\delta_3^0$ nanowire ($5{.}817\,{\rm eV} / {\rm atom}$) while among non-flat structures---to the $\kappa_8^1$ nanowire ($5{.}840\,{\rm eV} / {\rm atom}$). Boron can also form nanotubes and the open shape of $\kappa_8^1$ suggests it as well. The value of $E_{\rm b}$ for $\kappa_8^1$ is close to that of the buckled triangular boron sheet (see Fig.~\ref{fig6}).

The values of $E_{\rm b}$ for 1D structures are usually lower than those for 2D structures of the same material. It is worth noting that $E_{\rm b}$ for some of the obtained boron nanowires is larger than that for 2D boron triangular flat sheet ($s_6$-ft) \cite{GonzalezSzwacki2007b}. This is shown in Fig.~\ref{fig6} where we plot $E_{\rm b}$ for each structure from Fig.~\ref{fig5} as a function of $N_{\rm a}$. Figure~\ref{fig6} also shows three insets presenting the average number of generations across 5 computational series for each $N_{\rm a}$ value, the respective average computational time of GA-evolution on a single CPU \cite{CPU} machine, and normalized energy difference $\Delta E_{\rm b} / E_{\rm b,\ final}$ as a function of the generation number during GA-evolution resulting in the most energetically favorable structures, $\delta_3^0$ and $\kappa_8^1$. For both flat and non-flat structures, the number of generations needed to obtain the results increases with the number of atoms in the unit cell. On the other hand, at the logarithmic scale, the average computational time increases linearly with $N_{\rm a}$ for flat structures and rises rapidly for non-flat structures.

\section{Summary}
In summary, we study the structure of ultrathin boron 1D periodic structures using a DFT-based GA approach.  Based on our simulations, we have identified four main groups of structures: stripes with triangular or triangular and “square” motifs, stripes with larger holes, nanowires with an open tubular shape, and regular nanowires. The most stable thin structures are fully planar stripes with boron triangular motifs.
However, thicker stripes tend to form tubular-like structures that can be viewed as precursors of nanotubes or nanowires. Finally, several of the studied 1D structures are more stable than the boron hexagonal (triangular) sheet which gives hope that such thin structures may be experimentally obtained. Although the whole methodology developed in this work is tested on boron 1D structures, we believe that it has general predictive capabilities and can be used to study e.g. nanowires made of silicon or other monoatomic or polyatomic materials.

\section*{Acknowledgments}
This work is a result of the projects funded by the National Science Centre of Poland (Twardowskiego 16, PL-30312 Krak\'ow, Poland, \url{http://www.ncn.gov.pl/}) under the grants number \mbox{UMO-2013}\allowbreak{}\mbox{/11}\allowbreak{}\mbox{/B}\allowbreak{}\mbox{/ST3}\allowbreak{}\mbox{/04273} and \mbox{UMO-2016}\allowbreak{}\mbox{/23}\allowbreak{}\mbox{/B}\allowbreak{}\mbox{/ST3}\allowbreak{}\mbox{/03575}. 

\bibliographystyle{elsarticle-num}
\bibliography{v1}

\end{document}